\documentclass[10pt,conference]{IEEEtran}
\IEEEoverridecommandlockouts

\usepackage[numbers]{natbib}

\let\textcite\cite
\usepackage{amsmath,amssymb,amsfonts}
\usepackage{graphicx}
\usepackage{textcomp}
\usepackage{xcolor}
\usepackage{url}
\usepackage{algorithm}
\usepackage{algpseudocode}
\usepackage[hidelinks]{hyperref}
\usepackage[nameinlink,capitalize]{cleveref}
\usepackage{listings}

\makeatletter
\lst@Key{countblanklines}{true}[t]%
    {\lstKV@SetIf{#1}\lst@ifcountblanklines}

\lst@AddToHook{OnEmptyLine}{%
    \lst@ifnumberblanklines\else%
       \lst@ifcountblanklines\else%
         \advance\c@lstnumber-\@ne\relax%
       \fi%
    \fi}
\makeatother

\definecolor{codeblue}{rgb}{0.13,0.13,1}
\definecolor{codegreen}{rgb}{0,0.5,0}
\colorlet{numb}{magenta!60!black}

\definecolor{codegreen}{rgb}{0,0.6,0}
\definecolor{codegray}{rgb}{0.5,0.5,0.5}
\definecolor{codepurple}{rgb}{0.58,0,0.82}
\definecolor{backcolour}{rgb}{0.95,0.95,0.92}
\lstdefinestyle{mystyle}{
    commentstyle=\color{codegray},
    keywordstyle=\color{codeblue},
    numberstyle=\tiny\color{codegray},
    stringstyle=\color{codegreen},
    basicstyle=\ttfamily\small,
    breakatwhitespace=false,         
    breaklines=true,                 
    captionpos=b,                    
    keepspaces=true,                 
    numbers=left,                    
    numbersep=5pt,                  
    showspaces=false,                
    showstringspaces=false,
    showtabs=false,                  
    tabsize=2,
    breakindent=0pt,
}
\lstdefinelanguage{json}{
    basicstyle=\normalfont\ttfamily,
    keywordstyle=\color{codeblue},
    stringstyle=\color{codegreen},
    numbers=left,
    numberstyle=\scriptsize,
    stepnumber=1,
    numbersep=4pt,
    keepspaces=true,
    showstringspaces=false,
    tabsize=1,
    breaklines=true,
    string=[s]{"}{"},
    literate=
        *{0}{{{\color{numb}0}}}{1}
         {1}{{{\color{numb}1}}}{1}
         {2}{{{\color{numb}2}}}{1}
         {3}{{{\color{numb}3}}}{1}
         {4}{{{\color{numb}4}}}{1}
         {5}{{{\color{numb}5}}}{1}
         {6}{{{\color{numb}6}}}{1}
         {7}{{{\color{numb}7}}}{1}
         {8}{{{\color{numb}8}}}{1}
         {9}{{{\color{numb}9}}}{1}
         {true}{{\color{numb}true}}{3}
         {false}{{\color{numb}false}}{3}
}

\begin{document}

\title{pyMethods2Test: A Dataset of Python Tests Mapped to Focal Methods}

\author{\IEEEauthorblockN{Idriss Abdelmadjid}
\IEEEauthorblockA{\textit{University of Nebraska--Lincoln}\\
Lincoln, NE, USA \\
iabdelmadjid2@huskers.unl.edu}
\and
\IEEEauthorblockN{Robert Dyer}
\IEEEauthorblockA{\textit{University of Nebraska--Lincoln}\\
Lincoln, NE, USA \\
rdyer@unl.edu}
}

\maketitle

\begin{abstract}
Python is one of the fastest-growing programming languages and currently ranks as the top language in many lists, even recently overtaking JavaScript as the top language on GitHub.  Given its importance in data science and machine learning, it is imperative to be able to effectively train LLMs to generate good unit test cases for Python code.  This motivates the need for a large dataset to provide training and testing data.  To date, while other large datasets exist for languages like Java, none publicly exist for Python.  Python poses difficult challenges in generating such a dataset, due to its less rigid naming requirements.  In this work, we consider two commonly used Python unit testing frameworks: Pytest and unittest.  We analyze a large corpus of over 88K open-source GitHub projects utilizing these testing frameworks.  Using a carefully designed set of heuristics, we are able to locate over 22 million test methods.  We then analyze the test and non-test code and map individual unit tests to the focal method being tested.  This provides an explicit traceability link from the test to the tested method.  Our pyMethods2Test dataset contains over 2 million of these focal method mappings, as well as the ability to generate useful context for input to LLMs.  The pyMethods2Test dataset is publicly available on Zenodo at: \url{https://doi.org/10.5281/zenodo.14264518}
\end{abstract}


\section{Introduction}

Software testing has widely been acknowledged as an important activity for quickly developing secure and performant software.  Software tests provide developers the ability to detect more bugs, often at an earlier stage, and aid them in working toward fixes for those bugs.  Yet, despite their known benefits, often testing is one of the last things developers focus on as they instead prefer adding new features to the system~\cite{daka14}.

To help aid developers in creating tests, previous research has looked at various issues with the testing process such as the existence of code smells in test code~\cite{testsmells} and various issues with the maintenance and evolution of tests~\cite{pinto12}.  A lot of research has focused on automatically generating tests for developers, such as EvoSuite~\cite{evosuite}, Randoop~\cite{randoop}, and whole test suite generation~\cite{fraser13}.

Recently, large language models (LLMs) have shown great promise at generating text, including the generation of code.  Researchers have shown that such models are able to generate code, documentation~\cite{dvivedi24}, and even tests~\cite{llmtest}.  They also excel at specific software engineering activities, such as performing code reviews~\cite{lu23} and debugging~\cite{majdoub24}.  In order for these tasks to work, they must train the models on a sufficient number of examples.

Prior work has created large datasets for the use in training LLMs on test generation tasks.  For example, the Methods2Test dataset~\cite{tufano2022methods2test} contains a large number of test methods mapped to their focal methods for Java code.  Their work focuses on JUnit tests.  While such a dataset is immensely useful, its focus on Java limits how well the LLMs are able to train on generating tests for other languages, such as Python.  Python is one of the fastest growing languages~\cite{tiobe} and currently the most popular language on GitHub~\cite{octoverse}.

In this work, we create a dataset named pyMethods2Test.  This dataset provides a large number of open-source unit test methods and their mapping back to their focal methods (the main method they are testing).  We focus on Python code since it is so popular.  No prior work has generated a large dataset of Python test code mapped to their focal methods.  We mine a large number of repositories, looking for uses of two common Python testing frameworks: Pytest and unittest.  We then extract data about the tests, including file paths, line/column within the file, and names of the test methods.

Unlike languages like Java, where certain naming conventions are enforced such as the fact the filename must match the class name, Python tends to have less strict naming conventions.  The studied unit testing frameworks do still impose enough consistency that we are able to automatically map test methods to their focal methods in a large number of cases.  Our dataset contains over 22 million unit test methods with 10\% successfully mapped to their focal methods.  We provide this data in an easy-to-consume JSON format with additional metadata such as the repository, file paths, method locations, imported libraries, and source code module names.

Since one of the intended goals of this dataset is to provide training input for LLMs, we also provide the ability to generate a useful context for each data point.  For example, the context includes the focal method body as well as the class name, signatures of constructors and other methods, and attributes in the class.  Such context will aid LLMs in their ability to generate the mapped test methods.

The dataset and scripts used to build it are available on Zenodo~\cite{dataset}.

\section{Dataset}

To build this dataset, we mine a very large dataset of 2,661,596 GitHub repositories that indicated Python as the main programming language.  These repositories represent the raw data used as input to build other datasets for the Boa infrastructure~\cite{boa}.  They were cloned from GitHub after using the GitHub API to discover Python projects, with a preference for projects with higher star counts.  From this large set of projects, we then randomly selected projects for analysis with a goal of having around 100k projects in the final dataset.  The final dataset includes data from 88,846 Python projects.

To analyze the source files, we utilize the \texttt{ast} module from Python's standard library to parse the files into abstract syntax trees (ASTs).  We use this to find information about tests, the focal method, as well as to build the focal context.

All generated data is stored in JSON format.  The file structure is the name of the GitHub repository (e.g., \texttt{data/boalang/compiler/}) with all JSON data for that repository in this folder.  The filenames all include the commit hash representing the HEAD at the time of processing.  For each repository, there are three files in total, but two are considered intermediaries.  For verification and replication, we provide the intermediary files in a separate \texttt{raw-data.zip}~\cite{dataset}, as they are quite large (around 42GB, compared to less than 2GB for the focal data).

\begin{table}[ht]
    \centering
    \caption{Dataset Statistics}
    \label{tab:stats}
    \begin{tabular}{lr}
        \textbf{Repositories}  &      88,846 \\
        \textbf{All Files}     &  18,517,737 \\
        \textbf{Test Files}    &   1,289,630 \\
        \textbf{All Classes}   &  36,222,490 \\
        \textbf{All Methods}   & 222,020,293 \\
        \textbf{Test Methods}  &  22,662,037 \\
        \textbf{Focal Methods} &   2,198,378 \\
    \end{tabular}
\end{table}

First, we analyze each repository to collect information about defined classes and methods (note: in this paper, we use the phrase "methods" to denote either methods in a class or stand-alone functions) in each Python source file (\texttt{*.py}) along with their line numbers and indentation.  We also collect the module name of each file, which we approximate by looking for Python packages (\texttt{\_\_init\_\_.py} files) and appending the filename without an extension.  As shown in \Cref{tab:stats}, we located over 36M classes and 222M methods.  We store this data in JSON files with filenames such as \texttt{hash.json}.  Next, we use this information to locate all tests.

\subsection{Locating Test Files and Methods}

In the next stage, we identify test files, classes, and methods.  We focus on two popular Python testing frameworks: Pytest and the standard library's unittest.  We leverage the \texttt{import} statements in each file to determine if they use either library.  If either is used, we consider the file to be a test file and classify it as being either \texttt{pytest}, \texttt{unittest}, or \texttt{both} (as some code uses both libraries).  We then look at the filename, as Pytest has a default test discovery mechanism that is based on filename patterns.  Pytest requires files start with ``test\_'' or end with ``\_test'' while unittest is more relaxed (though many projects still follow similar naming conventions).  In total, we found over 1M matching test files.

We then analyze each discovered test file to locate possible test methods.  Here the strategy varies based on the framework.  For Pytest, we simply look for methods prefixed with ``test''.  For unittest, there must be a class declared that inherits from \texttt{unittest.TestCase}.  Within that class, the methods must be prefixed with ``test''.  For all passing methods, we collect their name, testing framework, file location, a list of imports that are non-test modules in the project, and called methods.  In total, we found over 22M test methods (10\% of all methods).

In order to eventually map a test method to a focal method, we first need to collect all methods invoked from the body of the test method.  We collect each method and store the set of called methods.  When collecting the called methods, we differentiate between methods in the project versus library methods.  To do this, we use the collected information about all classes and methods and inspect the imports for the test file.  If the test file is importing a module, class, or method from that data then we keep it.  This effectively filters out standard library calls (e.g., \texttt{len()}) as well as most external libraries such as Pandas.

\subsection{Locating Focal Files}

To locate the focal file, given a test method, we first look at the imports in the test file.  If only one non-test local file is being imported, we consider it the focal file.  Otherwise, if we do not find a direct match from the imports, we utilize the filenames.  First, we look at if a source file is a suffix of the test file's name.  For example, developers will often test a module \texttt{foo.py} with a file named \texttt{test\_foo.py}.  If we find exactly one match, we consider it the focal file.  If there are several matches, then we use a fuzzy string matching library and take the highest-scoring match.

\subsection{Locating Focal Classes}

Once we locate a focal file, we attempt to locate the focal class.  In Python, methods could be in a class or declared globally outside of a class.  Thus not all methods will map to a focal class.  We attempt to match the focal class first based on name and then fall back to using position-based information by walking back up the AST.

\subsection{Locating the Focal Method}

To locate the focal method, we consider all method invocations in the test method to be a non-library method.  We then check if the test's method name ends with the invoked method's name.  If there is no match, then we perform a fuzzy string match taking the best matching result while using a cutoff of 50 (based on a manual inspection).

\subsection{Focal Method Data Schema}

For each repository, we generate a JSON file containing the mapped focal method data.  An example of such a mapping is shown in \Cref{fig:schema}.  The files are a dictionary, where the keys are test file paths and the values are mapped data for that test file.  We include the identified focal file in \texttt{focal\_file} as well as a dictionary of all identified test \texttt{methods}.  For each test method, we include position information, the identified \texttt{focal\_class} (if any), and the identified \texttt{focal\_method} containing its own position information and the name of the method.

\begin{figure}[ht]
\begin{lstlisting}[language=json]
{
	"tests/unit/metrics/test_ffwd.py": {
		"focal_file": "gordon/metrics/ffwd.py",
		"methods": {
			"test_ffwd_protocol_connection_made":
			{
				"line": 23,
				"line_end": 32,
				"indent": 0,
				"focal_class": "gordon.metrics.ffwd.UDPClientProtocol",
				"focal_method": {
					"line": 59,
					"line_end": 67,
					"indent": 4,
					"name": "connection_made"
				}
			},
	...
\end{lstlisting}
\caption{Example of focal JSON data}
\label{fig:schema}
\end{figure}

\begin{figure*}[ht]
\begin{lstlisting}[language=python]
class LogRelay():                                                     # focal class
	def _create_metric(self, metric_name, value, context, **kwargs):    # focal method
		context = context or {}
		return {'metric_name': metric_name, 'value': value, 'context': context}

	def __init__(self, config): ...                                     # constructor
	def incr(self, metric_name, value=1, context=None, **kwargs): ...   # methods
	def timer(self, metric_name, context=None, **kwargs): ...
	def set(self, metric_name, value, context=None, **kwargs): ...
	def cleanup(self): ...

	                                                                    # class attributes (none)
	self.time_unit = config.get('time_unit', 1)                         # instance attributes
	self.logger = LoggerAdapter(level)
	self.counters = collections.defaultdict(int)
\end{lstlisting}
\caption{Example focal context from \texttt{spotify/gordon}}
\label{fig:context}
\end{figure*}

More specifically, the fields included are:

\begin{itemize}
    \item \texttt{test\_file}: The file path of the test file.
    \item \texttt{focal\_file}: The file path of the focal file (the file being tested).
    \item \texttt{methods}: A dictionary, where keys are test method names and values are mapped data for that test.
    \item \texttt{line}: The starting line of the method.
    \item \texttt{line\_end}: The ending line of the method.
    \item \texttt{indent}: The indentation level of the method, which helps indicate its position in the code structure.
    \item \texttt{focal\_class}: The class to which the tested method belongs, if applicable.
    \item \texttt{name}: The name of the original method under test.
\end{itemize}

\subsection{Context for the Focal Method}

Machine learning-based tools, such as large language models (LLMs), often need more than just a method in order to be accurate.  LLMs in particular support providing additional context with the prompt, and such context improves the accuracy of the generated results.  In this section, we discuss how we provide additional context for focal methods in our dataset.

All focal data is stored in files that follow a similar naming pattern: \texttt{user/repo/hash.focal.json}, where ``user/repo'' is the project name on GitHub and ``hash'' is the commit hash to analyze.  The commit hash ensures that the focal context, including method line numbers, remains accurate for the specific commit, as later commits might move methods around.

We provide a script to generate the focal context for a given repository, that expects such a file path as an argument.  The script assumes the repository is cloned in the ``repos'' directory or it will try cloning it.  It then generates the focal context.  An example of generated focal context is shown in \Cref{fig:context}.

If the focal method resides within a class, we follow a similar approach as \citet{tufano2020unit}, including the order we provide the various contexts.  Such context includes the following:

\begin{itemize}
    \item The focal class declaration.
    \item The focal method and body.
    \item The class constructor (\texttt{\_\_init\_\_}), as tests often will construct instances of this class.
    \item Additional method signatures: Any other methods in the class, as tests may often need to call these.
    \item Class and instance attributes.
\end{itemize}

\section{Applications}

The pyMethods2Test dataset bridges a gap by providing a large dataset of Python test methods mapped to their focal methods, where prior approaches focused on Java.  Such a dataset can enable many important applications for researchers, developers, and educators.

Researchers can mine the dataset to get a better idea of what kinds of testing strategies are actually employed in the wild, determine the popularity of Pytest versus unittest, mine the data for test smells~\cite{testsmells}, mine it for code style recommendations, mine for new test design patterns, etc.

In software development, the dataset supports the creation of tools for test automation, such as identifying test coverage gaps, optimizing workflows, and recommending additional test cases. Its detailed mappings and large dataset are well-suited for training LLMs and enabling predictive models that automate test-case generation, assist in fault localization, and reduce manual effort.

The dataset can aid educators by providing real-world examples of testing practices.  This can help students learn better testing techniques and be exposed to a wider range of testing styles.  They could also be used as examples in lectures or form the basis of evaluations such as homework or exams.

\section{Limitations}

One of the primary challenges of this study is the inherent variability in Python projects on GitHub. The lack of mostly standardized conventions for organizing test files and methods often complicates the process of accurately mapping test cases to their focal methods. Despite employing heuristics and string-matching techniques, some mappings may still be inaccurate, particularly in projects with unconventional naming schemes.

Another limitation is the focus on two primary testing frameworks, Pytest and Unittest. While these are widely used in the Python ecosystem, the exclusion of other, less popular, or custom testing frameworks may leave some testing practices unrepresented, potentially limiting the dataset's generalizability to projects using frameworks outside of Pytest and Unittest.

Additionally, the dataset assumes that methods with names similar to the test method and called within its body are the primary focal methods. While this assumption is valid in most cases, it may overlook scenarios where a test validates multiple methods or calls auxiliary methods as part of a broader testing context. The reliance on string matching for validation, while helpful, may also introduce ambiguity, particularly when method names are similar.

Despite these limitations, the dataset provides valuable insights and serves as a foundation for further research and improvement. Future iterations could address these challenges by expanding the range of testing frameworks, refining heuristics to handle more edge cases, and improving the handling of atypical naming schemes.

\section{Related Work}

Several prior works have addressed the challenge of test-to-code traceability, but many focus on languages other than Python or face limitations when applied to Python projects, especially those using Unittest or Pytest.

One such tool is Method2Test~\cite{tufano2022methods2test}, a Java-based solution that maps test cases to methods by leveraging Java’s structured naming conventions. While effective for Java projects using JUnit, its rigid structure contrasts with Python’s dynamic and more varied nature, making it less suitable for Python projects.

Another key example is TestRoutes~\cite{kicsi2020testroutes}, a manually curated dataset designed to map test cases to focal methods. However, TestRoutes mainly focuses on Java and faces scalability issues due to its manual curation, limiting its applicability for large-scale Python projects.

General-purpose static analysis tools, such as PyLint~\cite{pylint} and Bandit~\cite{bandit}, primarily focus on code quality and security rather than test-to-code traceability. While useful for identifying potential issues, they do not support mapping test cases to code.

Defects4J~\cite{defects4j} provides a set of reproducible bugs for Java, with tests to express the buggy behavior.  While this data is not directly mapping tests to focal methods, the buggy code the tests are mapped to could potentially lead to the focal method, especially if combined with an approach like SZZ~\cite{szz}.  This dataset however is substantially smaller than ours and focuses on Java while we focus on one of the most popular and fastest-growing languages Python.

While the goal is different, automated test generation approaches such as EvoSuite~\cite{evosuite} and A3Test~\cite{alagarsamy2024a3test} could be leveraged on a large dataset of projects to automatically create tests for a number of focal methods.  Such tests however would be generated and not hand-written by developers, meaning some downstream tasks, such as inferring code style, might not be able to utilize the data.

\section{Conclusion}

pyMethods2Test is the first Python dataset with more than 2 million methods mapped from almost 90k open-source projects, advancing software testing research by providing clear mappings between test and focal methods. The data is stored in JSON files for easy processing in downstream tasks.  We also provide the intermediate data, which includes information about all classes and methods in the studied repositories. We also provide a script to generate additional context for focal methods, useful for feeding into LLMs.  The scripts and data are available on Zenodo~\cite{dataset}.

\section*{Acknowledgment}
This work was supported in part by the U.S. National Science Foundation under grant CNS-2346327.

\clearpage
\bibliographystyle{IEEEtranN}
\bibliography{refs}

\end{document}